\documentclass[]{aipproc}
\layoutstyle{6x9}

\usepackage{times,colordvi,amsmath,epsfig,float,color,multicol,subfigure,natbib}
\usepackage[latin1]{inputenc}

\newcommand\hete{{\it HETE}}
\def\arcmin{$^{\prime}$}
\newcommand{\aaps}{{\it A\&AS}}
\newcommand{\apj}{{\it ApJ}}
\newcommand{\gcn}{{\it GCN}}
\def\arcsec{$^{\prime\prime}$}
\def\chandra{{\it Chandra}}
\def\lessim{\mathrel{\hbox{\rlap{\hbox{\lower4pt\hbox{$\sim$}}}\hbox{$<$}}}}
\def\gtrsim{\mathrel{\hbox{\rlap{\hbox{\lower4pt\hbox{$\sim$}}}\hbox{$>$}}}}

\begin{document}

\title
      [Observations of GRB~030528]
      {Chandra Observations of the Optically Dark GRB~030528}

\author{N. Butler}{
  address={Center for Space Research, Massachusetts Institute of Technology, MA},
}

\iftrue
\author{A. Dullighan}{
  address={Center for Space Research, Massachusetts Institute of Technology, MA},
}
\author{P. Ford}{
  address={Center for Space Research, Massachusetts Institute of Technology, MA},
}
\author{G. Ricker}{
  address={Center for Space Research, Massachusetts Institute of Technology, MA},
}
\author{R. Vanderspek}{
  address={Center for Space Research, Massachusetts Institute of Technology, MA},
}

\author{K. Hurley}{
  address={Space Sciences Laboratory, Berkeley, CA},
}
\author{J. Jernigan}{
  address={Space Sciences Laboratory, Berkeley, CA},
}

\author{D. Lamb}{
  address={Department of Astronomy, University of Chicago, IL},
}

\fi

\copyrightyear  {2003}

\begin{abstract}
The X-ray-rich GRB~030528 was detected by the HETE satellite
and its localization was rapidly disseminated.  However, early optical
observations
failed to detect a counterpart source.  In a 2-epoch ToO
observation with Chandra, we discovered a fading X-ray source
likely counterpart to GRB~030528.  The source brightness was typical
of X-ray afterglows observed at similar epochs.  Other observers
detected an IR source at a location consistent with the X-ray source. 
The X-ray spectrum is not consistent with a large absorbing column.
\end{abstract}

\maketitle

\section{Observations}
\label{sec:030528_dark}

The X-ray-rich (i.e. for the fluence $S$, 
$\log [S_X(2-30~{\rm kev})/S_\gamma(30-400~{\rm kev})]
> -0.5$)
GRB~030528 was detected by the \hete~satellite
\citep{atteia03} with a 2\arcmin~radius (90\% confidence)
SXC localization.  
The initial SXC error region was later revised
\citep{villasenor03} after the discovery of an
unaccounted-for systematic effect, resulting in a shift in position
center and an expansion of the error region to 2.5\arcmin~radius.
Early R-band observations reaching $R\approx 18.7$ roughly 
$140$ minutes after
the burst \citep{ayya03} and unfiltered observations
reaching 20.5 magnitude roughly 14 hours after the burst 
\citep{valentini03} failed to detect a counterpart.  

On 3 June, the {\it Chandra Observatory}~targeted the field of GRB~030528
as part of a series of GTO target-of-opportunity observations focusing
on optically-dark GRBs discovered by \hete.
The 25 ksec observation spanned the interval 12:22-20:08 UT, 5.97 - 
6.29 days after the burst. The revised SXC error region from 
\citet{villasenor03} was completely contained within the 
field-of-view of the \chandra~ACIS-S3 chip.
From 9 June 8:14 UT to 9 June 14:19 UT, 11.8 to 12.1 days post-burst,
\chandra~again 
targeted the field of GRB~030528 for a 20 ksec second epoch (E2) observation
with ACIS-S3.

\section{Chandra E1 Sources}

As reported in \citet{butler03a}, 4 candidate sources
were detected within the revised SXC error region.  Seven additional
non-stellar point sources were detected within the entire ACIS-S3
field-of-view.  Positions and other
data for these sources are shown in Table \ref{table:counts_030528}.
None of the sources
were anomalously bright relative to objects in \chandra~deep field
observations \citep[see, e.g.,][]{rosati02}.  We had performed
deep observations with Magellan prior to the \chandra~observation,
but none of the \chandra~sources were in our field of view.  However,
near-IR observations of a portion of the SXC error region containing two
of the E1 \chandra~sources revealed a fading Ks-band source \citep{greiner03}.
Between
0.7 and 3.6 days after the burst a fade by 0.9 magnitudes was observed
for a source spatially coincident with the brightest 
\chandra~source.
After the E1 \chandra~observation, deep observations in the
radio (6.8 days after the burst) \citep{frailnberg03} and in I-band 
(8.7 days after the burst, I$>$21.5) \citep{mirnhalp03}
failed to detect a counterpart source.

\begin{table}[ht]
\begin{tabular}{rlrcrcr}
\hline
 & & Epoch 1\, & &  Epoch 2\, & & \\
\# & Chandra Name & Net (Bg) & E2$_{\rm 90\%}$ &  Net (Bg) & $\Delta C$ & $P_C^{(\%)}$ \\
\hline
 1 & CXOU J170400.3-223710 & 39.5 (1.5) & 24.1,41.1 &  8.5 (2.5) & 6.97 & 0.01 \\
 4 & CXOU J170348.4-223826 & 30.1 (2.9) & 16.4,30.2 & 20.3 (2.7) & 0.01 & 37.7 \\
 9 & CXOU J170400.1-223548 & 10.8 (2.2) & 4.1,12.6  &  5.4 (3.6) & 0.17 & 28.8 \\
10 & CXOU J170354.0-223654 &  9.2 (2.8) & 3.4,12.6  &  8.3 (2.7) & 0.00 & 44.7 \\
... & & & & & & \\
 2 & CXOU J170358.7-224237 & 30.6 (3.4) & 14.1,28.0 & 23.9 (2.1) & 0.00 & 36.1 \\
 3 & CXOU J170355.7-223503 & 23.1 (2.9) & 11.8,24.6 & 15.1 (3.9) & 0.03 &  2.3 \\
 5 & CXOU J170342.8-223548 & 23.7 (4.3) & 12.4,26.0 & 10.7 (6.3) & 0.85 &  9.8 \\
 8 & CXOU J170403.9-223543 & 12.7 (2.3) & 5.7,16.0  &  4.4 (2.6) & 0.94 &  8.9  \\
14 & CXOU J170341.4-223646 &  6.1 (4.9) & 1.1,9.0   &  9.7 (5.3) & 0.00 & 48.8 \\
15 & CXOU J170411.2-224032 & 11.0 (4.0) & 4.3,14.4  &  5.5 (3.5) & 0.20 & 24.1 \\
17 & CXOU J170345.8-224133 & 10.5 (4.5) & 1.1,4.0   & ... & & \\
\hline
\end{tabular}
\caption{\noindent
\small
Four point sources are detected in the 0.5-8.0 keV band in the
\chandra~E1 observation
lying within the revised SXC error region.  Eight additional,
non-stellar point sources are detected in the ACIS-S3 field-of-view.
From the E1 net counts, we calculate E2$_{\rm 90\%}$,
the 90\% confidence
region for the expected net counts in E2, following
\citet{kraft99}.
The columns labeled ``$\Delta C$'' and ``$P_C$'' are explained in Section
\ref{sec:e2}.
Small values of $P_C$ indicate sources likely to have faded between E1 and E2.
Source \#17 was situated on a chip gap in E2.
}
\label{table:counts_030528}
\end{table}

\section{Afterglow Confirmed in E2}
\label{sec:e2}

Table \ref{table:counts_030528} shows the number of counts detected in 
E1, along with the 90\% confidence interval for E2 based on the E1 values.  
The E2 observations were reported in \citet{butler03b}. 
We have used a circular extraction region for each source, with radius set
to 2 times the 95\% encircled energy radius $r$.  This varies over the
chip and is approximated via $r=2.05-0.55*d+0.18*d^2$ arcsec, with $d$
measured in arcminutes from the center of the ACIS-S3 chip.  We use
an annular background region ten times larger than the signal region,
centered on and surrounding the signal region.  The exposure is 
calculated separately for each source extraction region in each epoch.

GRB X-ray afterglows typically fade in brightness with time as $t^{-1.3}$,
with $t$ measured from the GRB \citep{costa99}.  Assuming no spectral 
evolution, this implies a count-rate fade factor
of approximately 2.5 between E1 and E2.  We can test whether the data
for each source prefers a fade versus a constant count rate
by fitting the data for each source first with a single-rate model (Model A), 
then fitting the data with a model allowing the E2 rate to be lower than 
the E1 rate (Model B).  We do
the fits by maximizing the logarithm of the Poisson likelihood (i.e. the
\citet{cash79} statistic $C$).  We then simulate $10^4$ data sets for 
each source
using the count rate determined from Model A, and we count the number of these
which yield a larger $\Delta C$ than the observed value when fit with
Model B.  These fractions are expressed as probabilities ($P_C$) in Table 
\ref{table:counts_030528}.

Of the four sources in the revised
SXC error region (\#'s 1,4,9,10), source \#1 has faded far below
the 90\% confidence range established in E1.  The significance of the
fade is approximately 3$\sigma$, and we estimate the temporal
index (assuming a power-law fade) to be $\alpha=2.0 \pm 0.8$.  This
is somewhat steeper than the typical $t^{-1.3}$ fade for X-ray
afterglows \citep{costa99}, though it is characteristic of afterglows 
which have undergone a so-called ``jet-break'' \citep{frail01}.
None of the other \chandra~sources were observed to fade at a 
high level of significance, and source \#1 (also a fading IR source
as discussed above) is extremely likely to be counterpart to GRB~030528.

\section{Column Density Constraints}

\begin{figure}[ht]
\centering
\rotatebox{0}{\resizebox{15pc}{!}{\includegraphics{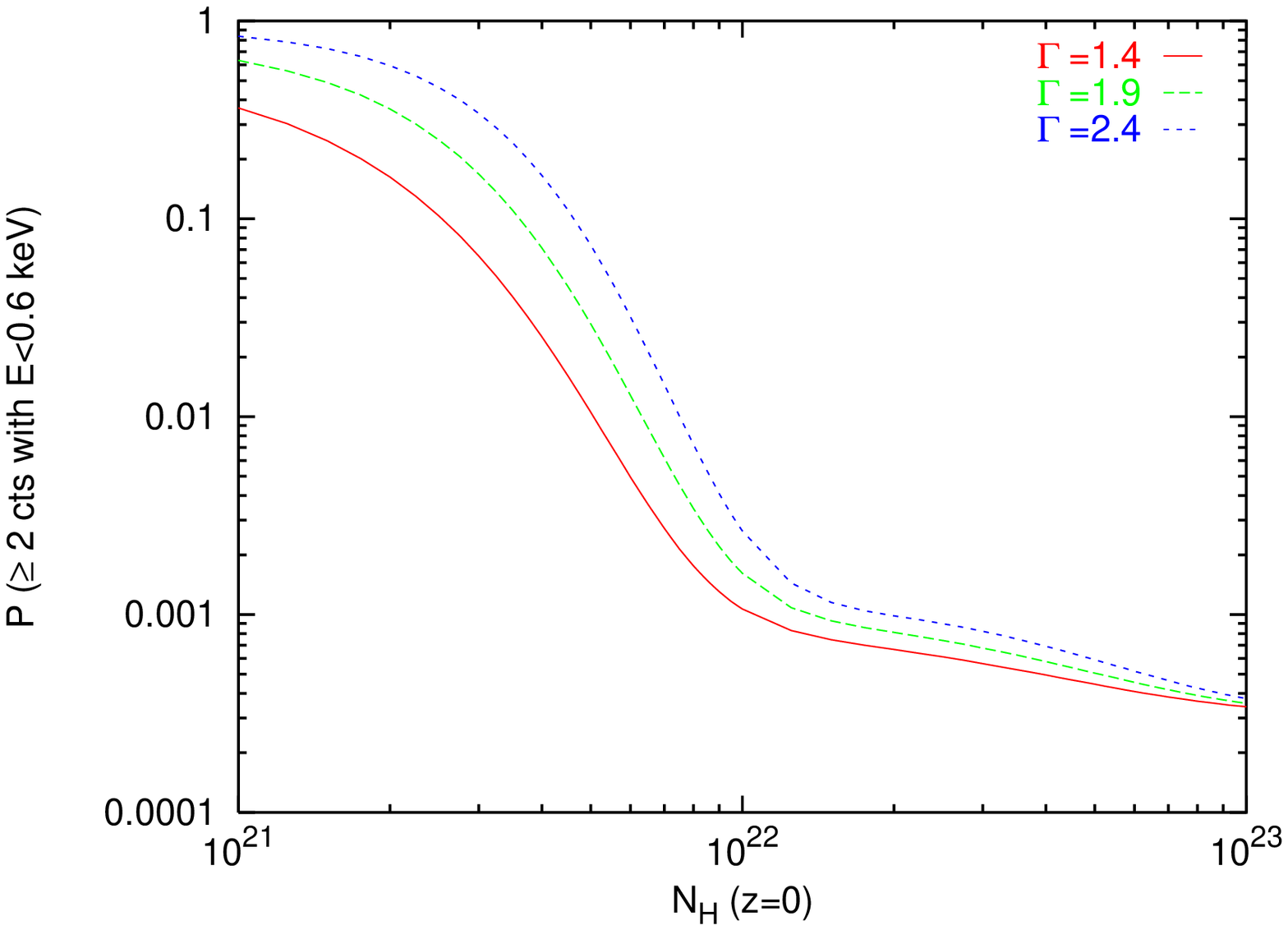}}}
\rotatebox{0}{\resizebox{15pc}{!}{\includegraphics{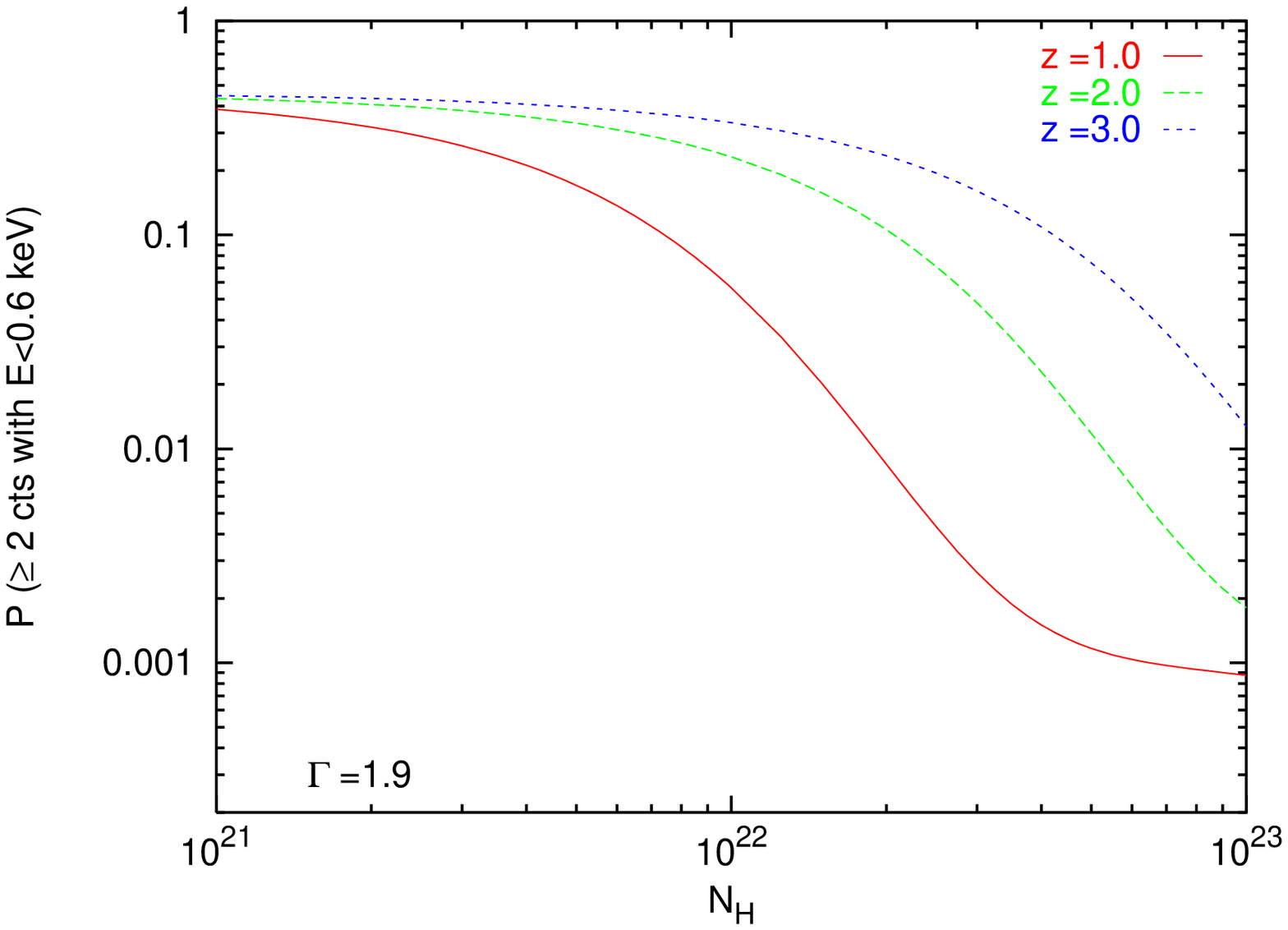}}}
\caption{
\small
The probability is very low that a power-law spectrum with the
indices shown (left plot) and $N_H\gtrsim 10^{22}$ cm$^{-2}$ could have
yielded
two source counts below 0.6 keV, as observed.  The possibility that
the counts could have come from the background is accounted for,
using an annular background
region approximately 100 times larger than the source region and
surrounding the source region.  The expected number of background counts
in the 3.3\arcsec radius source extraction region is 0.02 counts.
Because the host redshift is unknown, any local contribution in
excess of the Galactic column (right plot) may be less well constrained.
For $z=1$, it is likely that the local column is $\lessim 10^{22}$ cm$^{-2}$.
}
\label{fig:030528_nh}
\end{figure}

We reduce the spectral data using the standard 
CIAO\footnote{http://cxc.harvard.edu/ciao/} processing tools.
We use 
``contamarf''\footnote{http://space.mit.edu/CXC/analysis/ACIS\_Contam/script.html} to correct for the quantum efficiency degradation due to
contamination in the ACIS chips, important for energies below $\sim 1$ keV.
There are not enough source counts for detailed spectral fitting with the
\chandra~data.  However, we can use the instrumental response
determined in the steps above in combination with the total number
of detected source and background counts in the 0.5-8 keV band (Table
\ref{table:counts_030528}) and the number of source plus background
counts detected in the 0.5-0.6 keV band (2 counts) to constrain the 
model column density.  
Assuming a power-law spectrum, Figure \ref{fig:030528_nh} shows how
the two low-energy counts become increasingly improbable as the
column density increases.  
Except for the case of high redshift ($z>1$), Figure \ref{fig:030528_nh}
implies a column density $\lessim 10^{22}$ cm$^{-2}$.  This
implies an extinction in R-band of $A\lessim 3$ mag.  At high
redshift ($z\sim 3$), the column is likely $\lessim 10^{23}$ cm$^{-2}$.

\section{Conclusions}

An X-ray observation $\sim$6 days after GRB~030528 detected the afterglow
at a flux level ($1.4\times 10^{-14}$ erg cm$^{-2}$ s$^{-1}$)
typical for GRB X-ray afterglows \citep[see,][]{costa99} at that epoch.
Here we have assumed a typical power-law
spectrum with photon index $\Gamma=1.9$ and the Galactic 
$N_H=1.6\times 10^{21}$ cm$^{-2}$.  A second epoch 
observations decisively revealed that source \#1 had faded, establishing
securely that this was the counterpart X-ray afterglow to GRB~030528.
The X-ray spectrum appears to imply a fairly low column density,
which in turn implies a fairly low amount of reddening in the source
frame.  Thus, although the detection
of a near-IR counterpart with no detection of an optical counterpart
for this burst
perhaps points toward dust extinction, we find no supporting
evidence in the X-ray data.  The publication of additional photometric
data in various passbands for this burst, if available, would help
to constrain any possible extinction by dust in the GRB host galaxy.
We will perhaps learn that the afterglow to GRB~030528 was intrinsically
faint rather than heavily extincted, as appears to be common in may GRBs
\citep[see, e.g.,][]{berger02,ricker03}.

\end{document}